\documentclass{article}

\usepackage{arxiv}

\usepackage[utf8]{inputenc} % allow utf-8 input
\usepackage[T1]{fontenc}    % use 8-bit T1 fonts
\usepackage{hyperref}       % hyperlinks
\usepackage{url}            % simple URL typesetting
\usepackage{booktabs}       % professional-quality tables
\usepackage{amsfonts}       % blackboard math symbols
\usepackage{nicefrac}       % compact symbols for 1/2, etc.
\usepackage{microtype}      % microtypography
\usepackage{amsthm,dcolumn,graphicx,multicol,fancyhdr,bm}
\usepackage{amsmath}
\usepackage{multirow}
\usepackage{algorithm,algorithmic}
\usepackage{graphicx}
\usepackage{natbib}
\usepackage{doi}

\DeclareMathOperator*{\argmin}{arg\,min}

\title{Dynamic MRI reconstruction using low-rank plus sparse decomposition with smoothness regularization}

\author{Chee-Ming Ting\\
	School of Information Technology\\
        Monash University Malaysia\\
	\And
	Fuad Noman \\
        School of Information Technology\\
        Monash University Malaysia\\
        \And
        Rapha\"{e}l C.-W. Phan\\
        School of Information Technology\\
        Monash University Malaysia\\
        \And
        Hernando Ombao\\
        Statistics Program\\
        King Abdullah University of Science and Technology
}

\date{}

\renewcommand{\headeright}{}
\renewcommand{\undertitle}{}
\hypersetup{
pdftitle={A template for the arxiv style},
pdfsubject={q-bio.NC, q-bio.QM},
pdfauthor={David S.~Hippocampus, Elias D.~Striatum},
pdfkeywords={First keyword, Second keyword, More},
}

\begin{document}
\maketitle

\begin{abstract}
The low-rank plus sparse (L+S) decomposition model has enabled better reconstruction of dynamic  magnetic resonance imaging (dMRI) with separation into background (L) and dynamic (S) component. However, use of low-rank prior alone may not fully explain the slow variations or smoothness of the background part at the local scale. In this paper, we propose a smoothness-regularized L+S (SR-L+S) model for dMRI reconstruction from highly undersampled k-t-space data. We exploit joint low-rank and smooth priors on the background component of dMRI to better capture both its global and local temporal correlated structures. Extending the L+S formulation, the low-rank property is encoded by the nuclear norm, while the smoothness by a general $\ell_p$-norm penalty on the local differences of the columns of L. The additional smoothness regularizer can promote piecewise local consistency between neighboring frames. By smoothing out the noise and dynamic activities, it allows accurate recovery of the background part, and subsequently more robust dMRI reconstruction. Extensive experiments on multi-coil cardiac and synthetic data shows that the SR-L+S model outperforms several state-of-the-art methods in terms of recovery accuracy.
\end{abstract}

% keywords can be removed

\section{Introduction}
\label{sec:intro}

Dynamic magnetic resonance imaging (dMRI) is a powerful imaging modality widely used in clinical applications such as cardiac and perfusion imaging. It can reveal the spatial structure and temporal evolution of organs of interest. However, traditional dMRI is limited by slow imaging speed and low signal-to-noise ratio due to the additional temporal dimension and increased chance of imaging artifacts. The application of compressed sensing (CS) \cite{lustig2007sparse} combined with parallel MRI technique such as SENSE \cite{pruessmann1999} has enabled reconstruction of high-quality images from highly undersampled k-space data, significantly accelerating the dMRI process \cite{otazo2010}. To achieve undersampled MRI reconstruction, CS-based methods leverage on sparsity representation of images in original or transform domain, e.g., wavelets and temporal Fourier \cite{lustig2007sparse,lustig2006kt,otazo2010}.

In addition to the sparse prior of CS, low-rank models have also been explored for dMRI to characterize the temporal information correlation or redundancy across dynamic MR frames arising from slow changes of the same tissues. Earlier work on low-rank and sparse (L\&S) like kt-SLR \cite{lingala2011accelerated} assume image sequence as both low-rank and sparse. Alternatively, the low-rank plus sparse (L+S) decomposition is more natural for dynamic imaging because it separates the space-time data matrix ${\bf X} = {\bf L} + {\bf S}$ into a low-rank component (${\bf L}$) and a sparse component (${\bf S}$) that can represent the temporally-correlated background and dynamic foreground activities, respectively \cite{tremoulheac2014dynamic,otazo2015low}. By exploiting the inherent low-rankness of the slowly-changing background and stronger sparsity of images after the background subtracted, L+S has achieved better reconstruction quality than the CS-based methods. Various efficient algorithms have been developed to solve low-rank models for dMRI reconstruction. These include fast algorithms based on proximal gradient and variable splitting methods \cite{lin2018efficient}, and alternating gradient descent and minimization for MRI (altGDminMRI) \cite{Babu2023fast}. However, the low-rank regularization used in the above-mentioned studies can only describe the common low-dimensional subspace of the whole image sequence, but may not fully explain the slow variation or smoothness at the local range of the dynamic frames (e.g., smooth contrast changes in perfusion MRI).

In this paper, we propose a novel smoothness-regularized L+S (SR-L+S) framework for dynamic MRI reconstruction. Building on the L+S model, we leverage on both the low-rank and smoothness priors to simultaneously capture the global and local temporal correlations in the background components of the dMRI sequence. The reconstruction problem is formulated as a convex optimization, where nuclear norm and $\ell_1$ norm are used to, respectively, promote low-rankness and and sparsity on ${\bf L}$ and ${\bf S}$. We further incorporate a general smoothness regularizer that uses a $\ell_p$-norm penalty on the successive differences of the columns of the low-rank matrix ${\bf L}$ to enforce local similarity between neighbouring frames along the temporal dimension. The additional regularizer can provide further smoothing out of the random noise and foreground activities, thus enabling better recovery of the background component ${\bf L}$, and subsequently more robust image reconstruction. We consider both $\ell_2$ and $\ell_1$ smoothness regularization penalties, where the $\ell_1$ special case is an alternative formulation of the total variation (TV).

The main contributions of this work are as follows: (1) To our best knowledge, this is among the first work to incorporate temporal smoothness regularization into L+S reconstruction of dMRI. Although previous studies \cite{lingala2011accelerated,yao2018efficient} have combined low-rank and TV-based smoothness regularization for dMRI, they applied them on the reconstructed image ${\bf X}$ itself and not within the L+S framework as in the SR-L+S model that can facilitate simultaneous robust reconstruction and separation of dMRI. TV has also been used in L+S-based dMRI but only on the sparse component \cite{wang2020dynamic}. Moreover, these studies applied the TV regularization mainly in the spatial domain. In contrast, our model exploits the joint low-rank and smoothness priors for better modeling of the background component in the temporal domain to allow its accurate recovery from undersampled data. 
% (2) We develop an efficient algorithm based on proximal gradient method to solve the proposed SR-L+S formulation with closed-form updating equation for each involved variable. 
(2) We develop an efficient algorithm based on proximal gradient method to solve a convex optimization problem for SR-L+S model which consists of three regularization terms: nuclear norm for background part ${\bf L}$, $\ell_p$-norm for local smoothness in ${\bf L}$, and $\ell_1$-based sparsity for ${\bf S}$, with closed-form updating equation for each involved variable. 
(3) Experiments on cardiac perfusion and PINCAT synthetic data show significant improvement on recovery accuracy of the proposed model quantitatively and qualitatively over state-of-the-art low-rank methods.

\section{METHODS}
\label{sec:methods}

\subsection{L+S decomposition for dMRI reconstruction}

Consider a sequence of $N_t$ dynamic MR images of dimension $N_x \times N_y$. It is usually formulated as a space-time matrix (Casorati matrix) ${\bf X} = [{\bf x}_{1}, \ldots, {\bf x}_{N_t}] \in \mathbb{C}^{N_x N_y \times N_t}$ in which each column represents a vectorized temporal image or frame. A typical linear imaging model for dMRI can be written as
\begin{equation} \label{Eq:dMRI-model}
{\bf y} = {\bf E}({\bf X}) + {\bf n}
\end{equation}
where ${\bf y} \in \mathbb{C}^P$ is the undersampled ($k$-$t$)-space data measured during the acquisition, ${\bf E}: \mathbb{C}^{N_x N_y \times N_t} \rightarrow \mathbb{C}^P$ is the encoding or acquisition operator, and ${\bf n}$ is the noise vector. For parallel imaging with multiple receiver coils, the dimension of the k-space data is $P = N_s N_c$ with $N_s$ the total number of samples received from each receiver coil (across all frames) and $N_c$ is the number of coils, and ${\bf E} = {\bf F}_u {\bf S}_c$ where ${\bf F}_u$ is Fourier transform with undersampling and ${\bf S}_c$ denotes the coil sensitivities. The problem of image reconstruction is to recover the clean image ${\bf X}$ from the undersampled data ${\bf y}$, which is ill-posed under scenario $P \ll  N_x N_y \times N_t$ and requires regularization.

We consider the L+S framework \cite{otazo2015low, lin2018efficient} for reconstruction and separation of dMRI data, modeled as a superposition of a low-rank component ${\bf L}$ and a sparse component ${\bf S}$, ${\bf X} = {\bf L} + {\bf S}$. The low-rank matrix ${\bf L}$ represents the background component of images, which is assumed to change slowly over time or exhibit high correlation among frames. The ${\bf S}$ corresponds to the dynamic component which is rapidly changing. It captures the innovation in each frame in ${\bf X}$ with its background suppressed, which is already sparser that the original image, and has a much sparser representation under a proper sparsifying transform. The image reconstruction via L+S decomposition can be formulated as a convex minimization problem 
\begin{equation} \label{Eq:L+S}
% \{\widehat{\bf L}, \widehat{\bf S}\} = \argmin_{{\bf L}, {\bf S}} \frac{1}{2} \left\|{\bf E}({\bf L} + {\bf S}) - {\bf y} \right\|_2^{2} + \lambda_L\left\|{\bf L}\right\|_{*} +  \lambda_S \left\|\bf{T}({\bf S})\right\|_{1}
\min_{{\bf L}, {\bf S}} \frac{1}{2} \left\|{\bf E}({\bf L} + {\bf S}) - {\bf y} \right\|_2^{2} + \lambda_L\left\|{\bf L}\right\|_{*} +  \lambda_S \left\|\bf{T}({\bf S})\right\|_{1}
\end{equation}
where $\left\|{\bf L}\right\|_{*}$ is the nuclear norm or sum of singular values of ${\bf L}$, and $\left\|\bf{T}({\bf S})\right\|_{1}$ is the $\ell_1$ norm of transformed ${\bf S}$ under a sparsifying transform ${\bf T}$. As in \cite{otazo2015low, lin2018efficient}, we consider the Fourier transform operator along the temporal dimension for ${\bf T}$. The data fidelity term is captured by the $\ell_2$-norm term, the low-rankness of ${\bf L}$ is induced by the nuclear norm, and the sparsity of the transformed ${\bf S}$ by the $\ell_1$ penalty. The trade-off between these terms is controlled by the regularization parameters $\lambda_L$ and $\lambda_S$.

\subsection{The proposed model: L+S with smoothness}
The slowly-varying or possibly constant background components across frames may not be fully explained only by the low-rank prior in \eqref{Eq:L+S}. We exploit two types of prior knowledge jointly on the background component ${\bf L}$ for more robust image reconstruction: (1) \textit{Low-rankness} which considers that the entire temporal sequence in ${\bf L} = [{\bf l}_{1}, \ldots, {\bf l}_{N_t}]$ resides in a common low-dimensional subspace, revealing highly-correlated information at a global scale, and (2) \textit{Smoothness} which considers that neighbouring frames along the temporal dimension tend to vary smoothly, reflecting local information correlation or similarity at a relatively local scale. 

By jointly applying the local and global correlated priors, we propose a novel smoothness-regularized L+S (SR-L+S) model for dMRI reconstruction, which can be formulated as
\begin{equation} \label{Eq:SR-L+S}
\min_{{\bf L}, {\bf S}} \frac{1}{2} \left\|{\bf E}({\bf L} + {\bf S}) - {\bf y} \right\|_2^{2} + \lambda_L\left\|{\bf L}\right\|_{*} +  \lambda_S \left\|\bf{T}({\bf S})\right\|_{1} + \lambda_D\mathcal{R}({\bf L})
\end{equation}
where
\begin{equation}\label{Eq:smooth-prior}
% \mathcal{S}({\bf L}) = \sum_{t=2}^{N_t} \left( \left\|{\bf l}_{t} - {\bf l}_{t-1} \right\|_p \right)^{p}
\mathcal{R}({\bf L}) = \sum_{t=2}^{N_t} ( \left\|{\bf l}_{t} - {\bf l}_{t-1} \right\|_p )^{p}
\end{equation}
is smoothness regularizer controlled by parameter $\lambda_D$. Minimizing the $\ell_p$ ($p \leq 2$) norm on the differences of neighboring frames in \eqref{Eq:smooth-prior} can further enhance the temporal smoothness of the background component ${\bf L}$. We consider special cases $p=2$ and $p=1$. Let ${\bf D}$ be first-order finite difference matrix
\[
{\bf D} =
\left[
  \begin{array}{ccccc}
    -1 & 1 & 0 & \ldots & 0 \\
    0 & -1 & 1 & \ldots & 0 \\
    \vdots & \vdots  & \ddots & \ddots & \vdots \\
		0 & 0 & \ldots & -1 & 1 \\
  \end{array}
\right] \in \mathbb{R}^{(N_t-1) \times N_t}
\]
When $p=2$, \eqref{Eq:smooth-prior} simplifies to $\mathcal{R}({\bf L}) = \left\|{\bf L}{\bf D}^T \right\|_{F}^2 = \text{tr}({\bf L}{\bf D}^T{\bf D}{\bf L}^T)$ where $\left\|\cdot \right\|_{F}$ denotes the Frobenius norm. When $p=1$, $\mathcal{R}({\bf L}) = \left\| {\bf L}{\bf D}^T \right\|_{1}$ which is an alternative formulation of the popular temporal total variation (TV).

\subsection{Optimization}

We develop a proximal gradient method (PGM) to solve the optimization problem in \eqref{Eq:SR-L+S} for the proposed RS-L+S decomposition. Here, we present the algorithm for the $\ell_1$ smoothness prior, which can be applied similarly for the $\ell_2$ case. By introducing an auxiliary variable $\boldsymbol{\Phi} = {\bf L}{\bf D}^T \in \mathbb{C}^{N_x N_y \times N_t - 1}$, the problem is equivalent to
% \begin{small}
\begin{align}
\min_{{\bf L},{\bf S},\boldsymbol{\Phi}} & \frac{1}{2} \left\|{\bf E}({\bf L} + {\bf S}) - {\bf y}\right\|_2^{2} + \lambda_L \left\|{\bf L}\right\|_{*} + \lambda_S \left\|\bf{T}({\bf S})\right\|_{1} + \lambda_D \left\|\boldsymbol{\Phi}\right\|_{1} \notag \\ & s.t. \ \ \boldsymbol{\Phi} = {\bf L}{\bf D}^T \label{Eq:SR-L+S-phi}
\end{align}
% \end{small}
We further relax the equality constraint by transforming \eqref{Eq:SR-L+S-phi} to
\begin{align}
\min_{{\bf L},{\bf S},\boldsymbol{\Phi}} & \frac{\mu}{2} \left\|{\bf E}({\bf L} + {\bf S}) - {\bf y}\right\|_2^{2} + \frac{1}{2}  \left\|{\bf L}{\bf D}^T - \boldsymbol{\Phi} \right\|_F^{2} \notag  \\ & + \mu( \lambda_L \left\|{\bf L}\right\|_{*} + \lambda_S \left\|\bf{T}({\bf S})\right\|_{1} + \lambda_D \left\|\boldsymbol{\Phi}\right\|_{1}) \label{Eq:SR-L+S-phi-relax}
\end{align}
where $\mu > 0$ is a relaxation parameter.

By defining $\boldsymbol{\Theta} = ({\bf L};{\bf S};\boldsymbol{\Phi})$, the problem \eqref{Eq:SR-L+S-phi-relax} is a special case of a general convex problem of the form:
\begin{equation}\label{Eq:g+f}
\min_{\boldsymbol{\Theta}} F(\boldsymbol{\Theta}) = g(\boldsymbol{\Theta}) + f(\boldsymbol{\Theta})
\end{equation}
with
\begin{align}
g(\boldsymbol{\Theta}) & = \mu( \lambda_L \left\|{\bf L}\right\|_{*} + \lambda_S \left\|\bf{T}({\bf S})\right\|_{1} + \lambda_D \left\|\boldsymbol{\Phi}\right\|_{1}), \notag \\
f(\boldsymbol{\Theta}) & = \frac{\mu}{2} \left\|{\bf E}({\bf L} + {\bf S}) - {\bf y}\right\|_2^{2} + \frac{1}{2}  \left\|{\bf L}{\bf D}^T - \boldsymbol{\Phi} \right\|_F^{2} \notag
\end{align}
where $g$ is convex but not necessarily smooth, $f$ is smooth, convex and differentiable, and its gradient $\nabla f(\boldsymbol{\Theta})$ is Lipschitz continuous with constant $L_f>0$. The PGM forms a local approximation to the smooth term $f(\boldsymbol{\Theta})$ at a chosen point ${\boldsymbol{\Theta}}^k = ({\bf L}^k;{\bf S}^k;{\boldsymbol{\Phi}}^k)$, and updates $\boldsymbol{\Theta}$ iteratively using proximal operator
\begin{align}
\boldsymbol{\Theta}_{k+1} & = \argmin_{\boldsymbol{\Theta}} \ g(\boldsymbol{\Theta}) + \frac{L_f}{2} \| \boldsymbol{\Theta} - {\bf G}^k \|_{F}^2 \notag \\ \text{where} \ \ \   {\bf G}_k & = {\boldsymbol{\Theta}}_k - \frac{1}{L_f}\nabla f({\boldsymbol{\Theta}}_k) \label{Eq:proximal}
\end{align}

The problem \eqref{Eq:proximal} can be separable into the following independent subproblems:
\begin{align}
{\bf L}_{k+1} & = \argmin_{{\bf X}} \mu \lambda_L \left\|{\bf L}\right\|_{*} + \frac{L_f}{2} \| {\bf L} - {\bf G}^L_k \|_{F}^2 \label{Eq:updateX} \\
{\bf S}_{k+1} & = \argmin_{{\bf S}} \mu \lambda_S \left\|\bf{T}({\bf S})\right\|_{1} + \frac{L_f}{2} \| {\bf S} - {\bf G}^S_k \|_{F}^2 \label{Eq:updateS} \\
\boldsymbol{\Phi}_{k+1} & = \argmin_{\boldsymbol{\Phi}} \mu \lambda_D \left\|\boldsymbol{\Phi}\right\|_{1} + \frac{L_f}{2} \| \boldsymbol{\Phi} - {\bf G}^{\Phi}_k \|_{F}^2 \label{Eq:updatePhi}
\end{align}
where
\begin{small}
\begin{align}
{\bf G}^L_k & = {\bf L}_k - \frac{\mu}{L_f}( {\bf E}^H( {\bf E}({\bf L}_k + {\bf S}_k) - {\bf y})) - \frac{1}{L_f}({\bf L}_k {\bf D}^T {\bf D} - {\boldsymbol{\Phi}}_k{\bf D}) \notag \\
{\bf G}^S_k & = {\bf S}_k - \frac{\mu}{L_f}( {\bf E}^H( {\bf E}({\bf L}_k + {\bf S}_k) - {\bf y})) \notag \\
{\bf G}^{\Phi}_k & = {\boldsymbol{\Phi}}_k - \frac{1}{L_f} ({\boldsymbol{\Phi}}_k - {\bf L}_k{\bf D}^T) \notag.
\end{align}
\end{small}
% \begin{align}
% \begin{aligned}
% {\bf G}^L_k & = {\bf L}_k - \frac{\mu}{L_f}( {\bf E}^H( {\bf E}({\bf L}_k + {\bf S}_k) - {\bf y})) - \\
% & \hspace{4.5cm} \frac{1}{L_f}({\bf L}_k {\bf D}^T {\bf D} - {\boldsymbol{\Phi}}_k{\bf D}) \notag \\
% {\bf G}^S_k & = {\bf S}_k - \frac{\mu}{L_f}( {\bf E}^H( {\bf E}({\bf L}_k + {\bf S}_k) - {\bf y})) \notag \\
% {\bf G}^{\Phi}_k & = {\boldsymbol{\Phi}}_k - \frac{1}{L_f} ({\boldsymbol{\Phi}}_k - {\bf L}_k{\bf D}^T) \notag.
% \end{aligned}
% \end{align}

Here ${\bf E}^H$ denotes the adjoint operator of ${\bf E}$ which maps a vector to a matrix. Each of these subproblems has closed-form solutions using the proximal maps for nuclear norm or $\ell_1$ norm as follows:
\begin{align}
{\bf L}_{k+1} & = {SVT}_{{\mu \lambda_L}/{L_f}} \left[{\bf G}^L_k \right] \\ \label{Eq:solL}
{\bf S}_{k+1} & = {\bf T}^H(\mathcal{S}_{{\mu \lambda_S}/L_f} \left[ {\bf T}({\bf G}^S_k) \right]) \\ %\label{Eq:solS}
\boldsymbol{\Phi}_{k+1} & = \mathcal{S}_{{\mu \lambda_D}/{L_f}} \left[{\bf G}^{\Phi}_k\right] \label{Eq:solPhi}
\end{align}
where $SVT$ denotes the singular value thresholding operator for complex-valued numbers given by ${SVT}_\tau[{\bf Z}] = {\bf U}\mathcal{S}_\tau[\boldsymbol{\Sigma}]{\bf V}^H$ where ${\bf Z} = {\bf U}\boldsymbol{\Sigma}{\bf V}^H$ is the singular value decomposition (SVD) of a complex-valued matrix ${\bf Z}$, and $\mathcal{S}_{\tau}[{\bf Z}]$ denotes the element-wise soft-thresholding or shrinkage operator at threshold $\tau>0$, i.e., $\mathcal{S}_{\tau}[z] = \frac{z}{|z|}\max(|z|-\tau,0)$. Here ${\bf T}^H$ is adjoint operator of ${\bf T}$, defined by the corresponding inverse Fourier transform. We terminate the iteration when $\left\|{\bf L}_{k+1} + {\bf S}_{k+1} - ({\bf L}_{k} +{\bf S}_{k})\right\|_{F} \leq 10^{-5} \left\|{\bf L}_{k}+{\bf S}_{k}\right\|_{F}$. The proposed PGM method for solving the smoothness-regularized L+S decomposition and reconstruction of dMRI is summarized in Algorithm 1.

\begin{algorithm}[!t]
\caption{Proximal Gradient L+S with Smoothness}
\begin{algorithmic}[1]
\renewcommand{\algorithmicrequire}{\textbf{Input:}}
\renewcommand{\algorithmicensure}{\textbf{Output:}}
\REQUIRE Undersampled k-t data ${\bf y}$, data acquisition operator ${\bf E}$, temporal Fourier transform ${\bf T}$,   difference matrix ${\bf D}$, regularization parameters $\lambda_L$, $\lambda_S$, $\lambda_D$, $\mu$, $L_f$.
\renewcommand{\algorithmicrequire}{\textbf{Initialize:}}
\REQUIRE ${\bf M}_0 =  {\bf L}_0 = {\bf E}^H{\bf y}$, ${\bf S}_0 = {\bf 0}$, $\boldsymbol{\Phi}_0 = {\bf 0}$
\WHILE {not converged}
    % \STATE ${\bf L}_{k+1} = {SVT}_{\frac{\mu \lambda_L}{L_f}} \left[ {\bf L}_k - \frac{1}{L_f}(\mu{\bf M}_k + {\bf L}_k {\bf D}^T {\bf D} - {\boldsymbol{\Phi}}_k{\bf D}) \right]$
    \STATE ${\bf L}_{k+1} = {SVT}_{{\mu \lambda_L} / L_f} \left[ {\bf L}_k - \frac{1}{L_f}(\mu{\bf M}_k + {\bf L}_k {\bf D}^T {\bf D} - {\boldsymbol{\Phi}}_k{\bf D}) \right]$
    % \STATE ${\bf L}_{k+1} = {SVT}_{{\mu \lambda_L} / L_f} \left[ {\bf L}_k - \frac{1}{L_f}(\mu{\bf M}_k + \right.$\\
    %  $\hspace{4.8cm}\left.{\bf L}_k {\bf D}^T {\bf D} - {\boldsymbol{\Phi}}_k{\bf D}) \right]$
    \STATE ${\bf S}_{k+1} = {\bf T}^H(\mathcal{S}_{{\mu \lambda_S}/{L_f}} \left[ {\bf T}({\bf S}_k - \frac{\mu}{L_f}{\bf M}_k) \right])$
    \STATE $\boldsymbol{\Phi}_{k+1} = \mathcal{S}_{{\mu \lambda_D}/{L_f}} \left[ {\boldsymbol{\Phi}}_k - \frac{1}{L_f} ({\boldsymbol{\Phi}}_k - {\bf L}_k{\bf D}^T) \right]$
    \STATE ${\bf M}_{k+1} = {\bf E}^H( {\bf E}({\bf L}_{k+1} + {\bf S}_{k+1}) - {\bf y}))$
\ENDWHILE
\ENSURE ${\bf L}, {\bf S}, \boldsymbol{\Phi}$
\end{algorithmic}
\end{algorithm}

\begin{table*}[!ht]
	% \caption{Performance comparison of different methods.}
        \caption{Quantitative comparison of reconstruction using different methods on PINCAT and cardiac perfusion data.}
	\label{table:bigresult}
	\centering
	\resizebox{1\columnwidth}{!}{
	\begin{tabular}{llcccc}
		\hline \hline
		Dataset &  Method & SER  (dB)& PSNR (dB)& SSIM (n.u) & HFEN (n.u) \\ \hline
		\multirow{8}{*}{PINCAT}   
            &Zerofilled      &18.8056      &31.2209      &0.8069       &0.5108\\      
            &kt-SLR \cite{lingala2011accelerated}       &18.8078      &31.2230      &0.8071       &0.5108\\      
            &L+S \cite{otazo2015low}      &24.5159      &36.9767      &0.9594       &0.2513\\      
            &L+S-FISTA \cite{lin2018efficient}    &24.6016      &37.0687      &0.9597       &0.2465\\      
            &L+S-POGM \cite{lin2018efficient}       &24.5765      &37.0263      &0.9597       &0.2478\\
            &altGDminMRI1 \cite{Babu2023fast} &25.4515      &38.3994      &0.9292       &0.2378\\ 
            \cline{2-6}
            &SR-L+S-$\ell_2$ (ours)      &26.0206      &38.5896      &0.9437       &0.2146\\      
            &SR-L+S-$\ell_1$ (ours)     &\textbf{27.7572 }     &\textbf{40.2958}      &\textbf{0.9637}       &\textbf{0.1711}\\  
		  \hline
		\multirow{8}{*}{Cardiac} 
            &Zerofilled      &6.1757       &19.1928      &0.5037       &0.8448\\      
            &kt-SLR \cite{lingala2011accelerated}      &-1.9176      &10.9185      &0.0977       &1.3755\\      
            &L+S \cite{otazo2015low}        &13.3158      &26.2230      &0.7874       &0.4546\\      
            &L+S-FISTA \cite{lin2018efficient}    &13.3129      &26.2200      &0.7868       &0.4548\\      
            &L+S-POGM \cite{lin2018efficient}       &15.3863      &28.2842      &0.8509       &0.3243\\      
            &altGDminMRI1 \cite{Babu2023fast} &15.5309      &28.3784      &0.8678       &0.2547\\ 
            \cline{2-6}
            &SR-L+S-$\ell_2$ (ours)        &16.3301      &29.2059      &0.8739       &0.2715\\      
            &SR-L+S-$\ell_1$ (ours)     &\textbf{16.5811}      &\textbf{29.4659}      &\textbf{0.8787}       &\textbf{0.2526}\\       
            \hline \hline
	\end{tabular}}
\end{table*}

\section{Experimental Results}
\label{sec:results}
% To demonstrate the effectiveness of the proposed methods, we performed comparisons with several baseline and state-of-the-art methods, including Zerofilled (missing k-space data is replaced with zeros before applying the inverse Fourier transform), kt-SLR (utilizes sparsity in both spatial and temporal domains for joint iterative reconstruction) \cite{lingala2011accelerated}, ISTA (iterative algorithm that enforces sparsity in the image domain through soft thresholding) \cite{lin2018efficient}, FISTA (enhances upon traditional iterative reconstruction algorithms by incorporating acceleration techniques while enforcing sparsity through soft thresholding) \cite{lin2018efficient}, POGM (applies optimization algorithms that jointly promote sparsity in the image domain and fidelity  with improved computational efficiency) \cite{lin2018efficient}, and altGDminMRI1 (employs an alternate gradient descent approach combined with minimization techniques while enforcing sparsity and fidelity constraints) \cite{Babu2023fast}.

% We evaluate the effectiveness of the proposed model for dMRI reconstruction.

\subsection{Datasets}
% \subsubsection{Cardiac Perfusion Dataset}
% Two distinct datasets were used to assess the proposed methods performance, including
% The experiments are conducted on two dMRI datasets: 
We evaluate the effectiveness of the proposed SR-L+S model for dMRI reconstruction on two datasets. 1) the physiologically enhanced non-uniform cardiac torso (PINCAT) numerical phantom data used in \cite{lingala2011accelerated,lin2018efficient} and 2) in vivo cardiac perfusion MRI data from \cite{otazo2015low,lin2018efficient}. The groud-truth PINCAT data has spatial dimension $N_x \times N_y=128 \times 128$ with $N_t=50$ temporal frames. To emulate the multi-coil setting for the PINCAT data, we simulated sensitivity maps of 32 coils (4 rings of 8 coils) compressed to $N_c=8$ coils, following setup in \cite{lingala2011accelerated, lin2018efficient}.
% As in \cite{lin2018efficient, Babu2023fast} for PINCAT, we employed simulated coil sensitivity maps to facilitate algorithmic comparisons in a multi-coil context. The use of coil compression led to a reduced number of coils, specifically $N_c=8$.
The cardiac data consist of images with size $N_x \times N_y=128 \times 128$, with $N_t=40$ temporal frames and $N_c=12$ coils with a retrospective undersampling factor of 8 as in \cite{lingala2011accelerated, otazo2015low, lin2018efficient}.

% Cardiac: retorspective undersampled using Cartesian variable density random undersampling at reduction factors (R); R=8 \cite{Babu2023fast}
% Pincat: pseudo-radial undersampling mask Ω, i.e., a Cartesian trajectory that closely approximates a radial trajectory, with 24 spokes per frame, corresponding to a acceleration factor of 128/24 ≈ 5.3. We added zero mean Gaussian noisesuch that the signal to noise ratio is 46 dB. \cite{lin2018efficient}

% \subsection{Evaluation Metrics}
% \subsection{Quantitative Comparison with Existing Methods}
\subsection{Performance metrics \& comparison}
% To quantitatively assess the performance of the proposed methods in reconstructing MRI images compared to fully-sampled reference images, we utilized several widely used metrics. These metrics include signal error rate (SER), peak signal-to-noise ratio (PSNR), structural similarity index (SSIM), and high-frequency error norm (HFEN). For HFEN, we employed a LoG filter with a kernel size of $15 \times 15$ pixels and a standard deviation of $1.5$ pixels.
To quantitatively assess the dMRI reconstruction performance of each method, we utilized performance metrics commonly used in CS-based reconstructions. These include signal error rate (SER), peak signal-to-noise ratio (PSNR), structural similarity index (SSIM), and high-frequency error norm (HFEN) \cite{ravishankar2010mr}. The reported PSNR and SSIM are averages over all the frames. For HFEN, we employed a LoG filter with a kernel size of $15 \times 15$ pixels and a standard deviation of $1.5$ pixels. A higher SER, PSNR and SSIM and lower HFEN indicate better reconstruction respective to the original reference images. 

We compared the proposed SR-L+S models with 5 state-of-the-art low-rank methods for dMRI: kt-SLR \cite{lingala2011accelerated}, L+S with proximal gradient method \cite{otazo2015low}, L+S solved by fast iterative shrinkage-thresholding (FISTA) algorithm \cite{lin2018efficient}, L+S solved by proximal optimized gradient method (POGM) \cite{lin2018efficient}, and altGDminMRI1 \cite{Babu2023fast} which used an alternating gradient descent combined with minimization techniques assuming a hierarchical low-rank model. Zerofilled reconstruction \cite{bernstein2001effect} is also included as baseline.

% \subsection{Model configuration}
\subsection{Parameter settings}
% \subsection{Implementation details}
% We followed the author-provided parameter settings used in the original implementation of the competing methods on the cardiac and PINCAT data.
 % For the experiments in this study, we conducted empirical tuning of hyperparameters of the proposed models to achieve optimal results. The reported outcomes are based on specific parameter configurations tailored for each dataset.
 The regularization parameters were empirically selected on respective dataset by choosing an parameter set that gives optimal reconstruction performance over a range of values. 
 For the PINCAT dataset, the SR-L+S-$\ell_1$ used $\lambda_L$=0.01, $\lambda_S$=0.001, $L_f$=3, and $\lambda_D$=0.01; the SR-L+S-$\ell_2$ used $\lambda_L$=0.01, $\lambda_S$=0.001, $L_f$=5, and $\lambda_D$=0.01. For the Cardiac dataset, parameters $\lambda_L$=0.01, $\lambda_S$=0.01, $L_f$=3, and $\lambda_D$=0.005 is set for the SR-L+S-$\ell_1$, and $\lambda_L$=0.01, $\lambda_S$=0.01, $L_f$=5, and $\lambda_D$=0.005 for the SR-L+S-$\ell_2$. We set $\mu = 1$ for all experiments.
 % These parameter choices were made to optimize the performance of the proposed models on the respective datasets.
 For the competing methods, we followed the original parameter settings provided by the authors for the implementation on the PINCAT and cardiac data.

\begin{figure*}[!ht]
\centerline{\includegraphics[width=1\columnwidth]{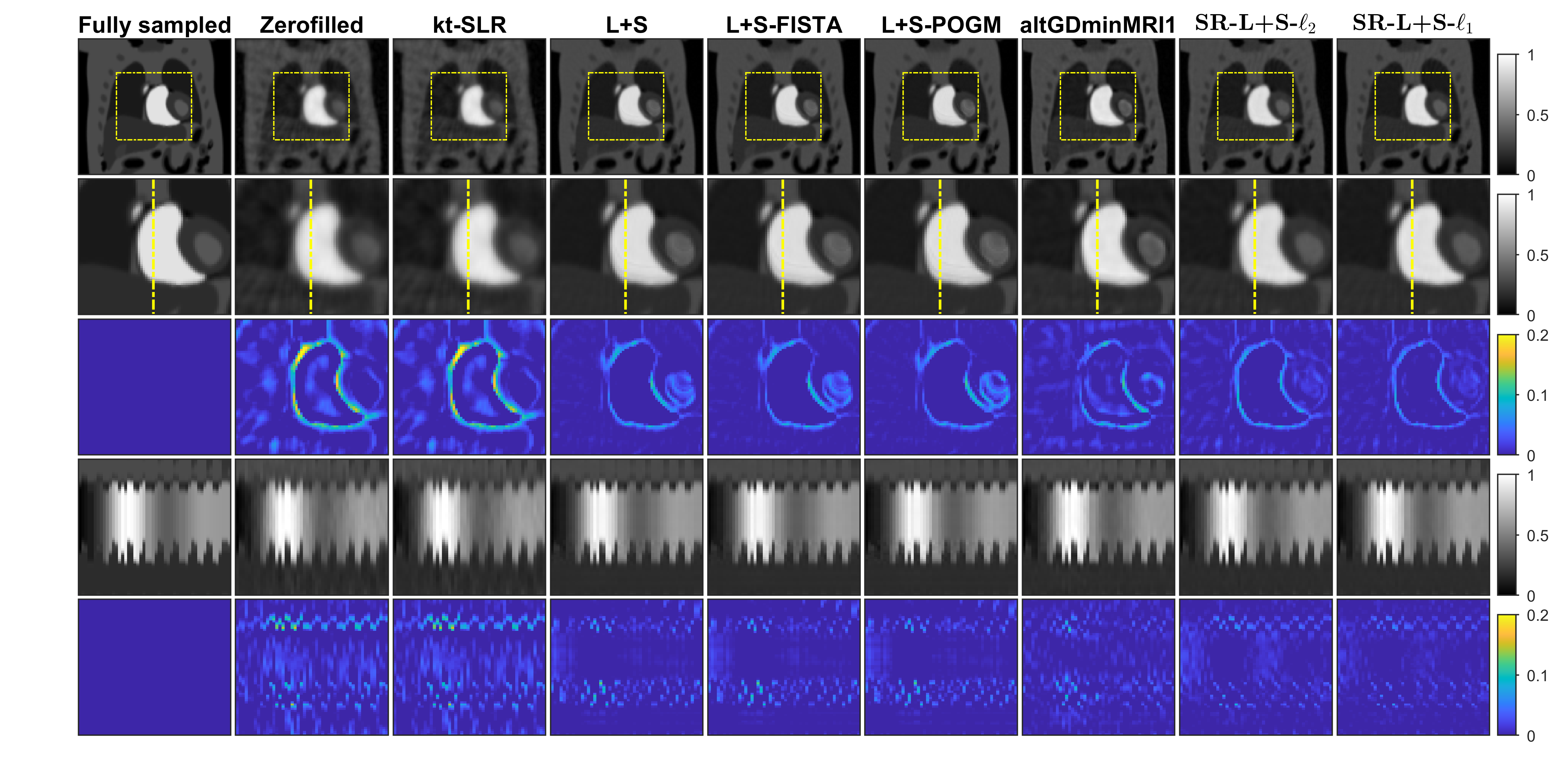}}
\vspace{-0.1in}
% \caption{Results on PINCAT dataset. Row 1 shows the full-view of an original fully-sampled frame (14th frame) and its reconstructions using different methods. In row 1, the enlarged views of the respective regions highlighted with yellow box in row 1. Row 3 displays reconstruction residual images, plotted on a scale of [0, 0.2]. Row 4 showcases time profile images with the selected cut line (slice) at the center ($x=64$). Row 5 presents corresponding time profile residual images.}
\caption{Comparison of reconstruction results of PINCAT synthetic data (14th frame) with 8-fold undersampling using different methods. First row: Original (fully-sampled) and reconstructed images. Second row: Enlarged views of yellow boxed region. Third row: Error maps (display scale of [0, 0.2]) with respective to the original image. Forth \& fifth rows: y-t images at the vertical cut line (slice $x=64$) in the reconstructed images, and the corresponding error maps, respectively.}
\label{fig:pincat_res}
\vspace{-0.1in}
\end{figure*}

\begin{figure*}[!ht]
\centerline{\includegraphics[width=1\columnwidth]{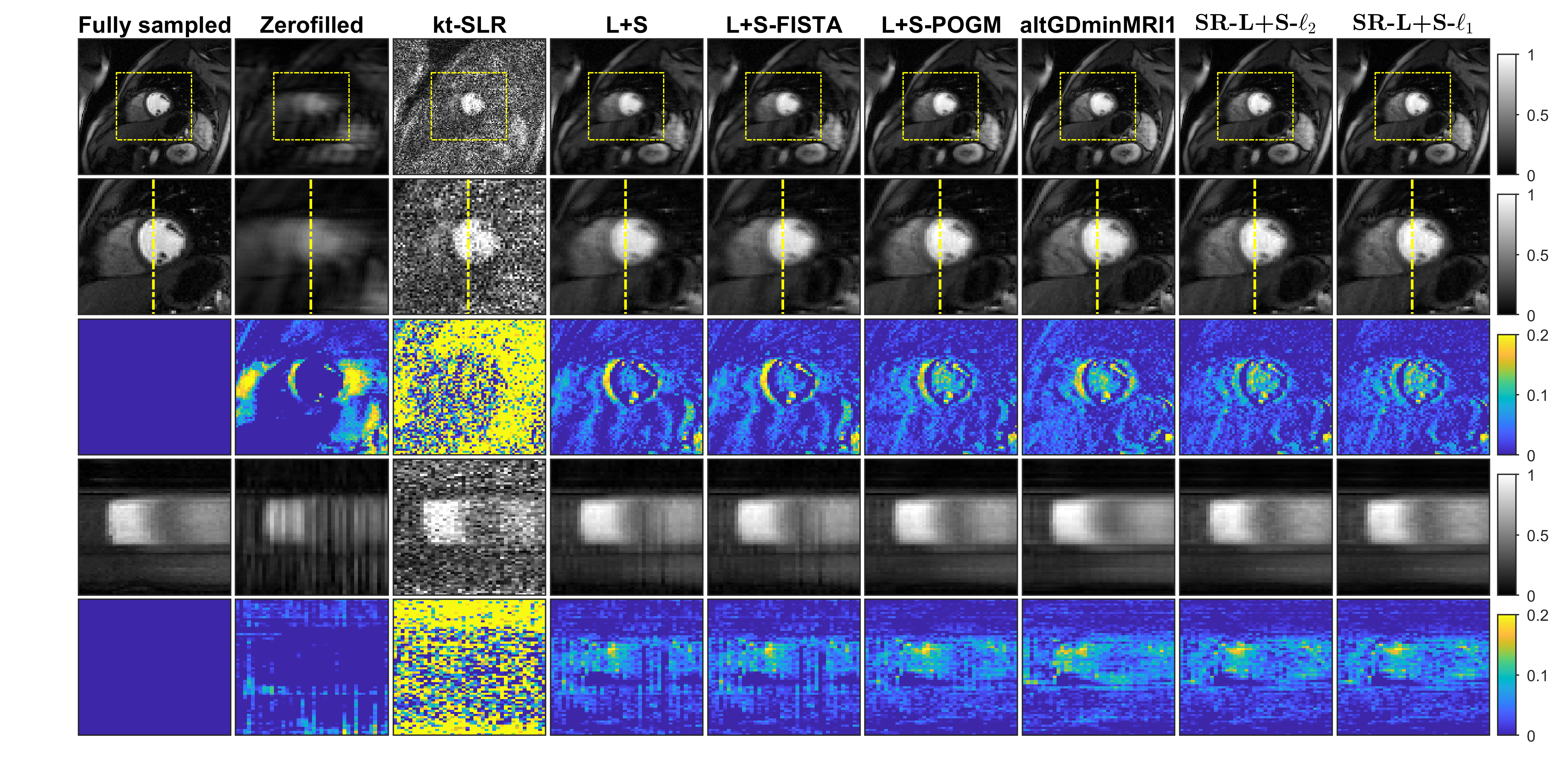}}
\vspace{-0.1in}
% \caption{Comparison of reconstruction results of multi-coil cardiac perfusion MRI with 8-fold undersampling using different methods. First row: Original (fully-sampled) and reconstructed images. Second row: Enlarged views of corresponding heart region of interests (ROI) with myocardial wall enhancement. Third row: Error maps with respective to the original image. Forth \& Fifth rows: X-t images at the horizontal cut line (slide $y=64$) in the reconstructed images, and the corresponding error maps, respectively.}
\caption{Comparison of reconstruction results of multi-coil cardiac perfusion MRI with 8-fold undersampling using different methods. First row: Original (fully-sampled) and reconstructed images. Second row: Enlarged views of corresponding heart region with myocardial wall enhancement. Third row: Error maps with respective to the original image. Forth \& fifth rows: y-t images at the vertical cut line (slice $x=64$) in the reconstructed images, and the corresponding error maps, respectively.}
\label{fig:perf_res}
\vspace{-0.1in}
\end{figure*}

\subsection{Quantitative \& qualitative results}

\subsubsection{Reconstruction accuracy}

Table~\ref{table:bigresult} shows the quantitative results of the proposed SR-L+S and comparison methods measured by SER, PSNR, SSIM and HFEN. It is obvious that SR-L+S models achieve the highest reconstruction accuracy compared to other methods, with substantial improvements of all evaluation metrics on both datasets. The highest SER, PSNR and SSIM, and lowest HFEN of our models indicate the minimal level of noise and distortion in the reconstructed images, while preserving the fine details and structural information of the original images. The results imply that the incorporation of smoothness regularization of the background components in SR-L+S can enhance dMRI reconstruction performance of the original L+S formulations. Among the SR-L+S models, use of $\ell_1$ smoothness prior outperforms $\ell_2$ smoothness, suggesting the advantage of piece-wise smoothness induced by the $\ell_1$-penalty. We can see that L+S models perform better that the L\&S model, with improved reconstruction by the use of efficient algorithms such as FISTA, POGM and altGDminMRI1.

\subsubsection{Visual results}

Fig.~\ref{fig:pincat_res} and Fig.~\ref{fig:perf_res} show the reconstruction results of different methods in the spatial and time domains on the PINCAT and cardiac perfusion data, respectively. The first and second rows show the reconstructed images (14th frame) and magnified views or a particular region of interest (ROI) in the yellow-box area. We can see that the proposed SR-L+S models provide visually higher-quality reconstruction on both datasets. They reconstruct cleaner images with lower aliasing artifacts and better preservation of sharp edges and fine details of the images. Blurry reconstructions are evident particularly for the kt-SLR and the original L+S models. The Kt-SLR also suffers from noise enhancement on the cardiac data as also reported in \cite{Babu2023fast}. The lower residual artifacts of our method can be indicated clearly in the error maps relative to the ground-truth images (third row). The reconstruction errors of the other methods are more noticeable that that of the proposed method, especially around the edges, e.g., of the papillary muscle for the cardiac data. The y-t images (forth row) and its corresponding error maps (fifth row) also show that our method can capture the dynamic information more accurately. The reconstructed time profiles of competing methods tend to over-smooth with more pronounced motion blurring especially on the cardiac data. These qualitative findings consistently indicate the more accurate and better reconstruction performance of our method , as corroborated by the quantitative measures in Table~\ref{table:bigresult}. 

\begin{figure}[t!]

\begin{minipage}[b]{1.0\linewidth}
  \centering
  \centerline{\includegraphics[width=8.5cm]{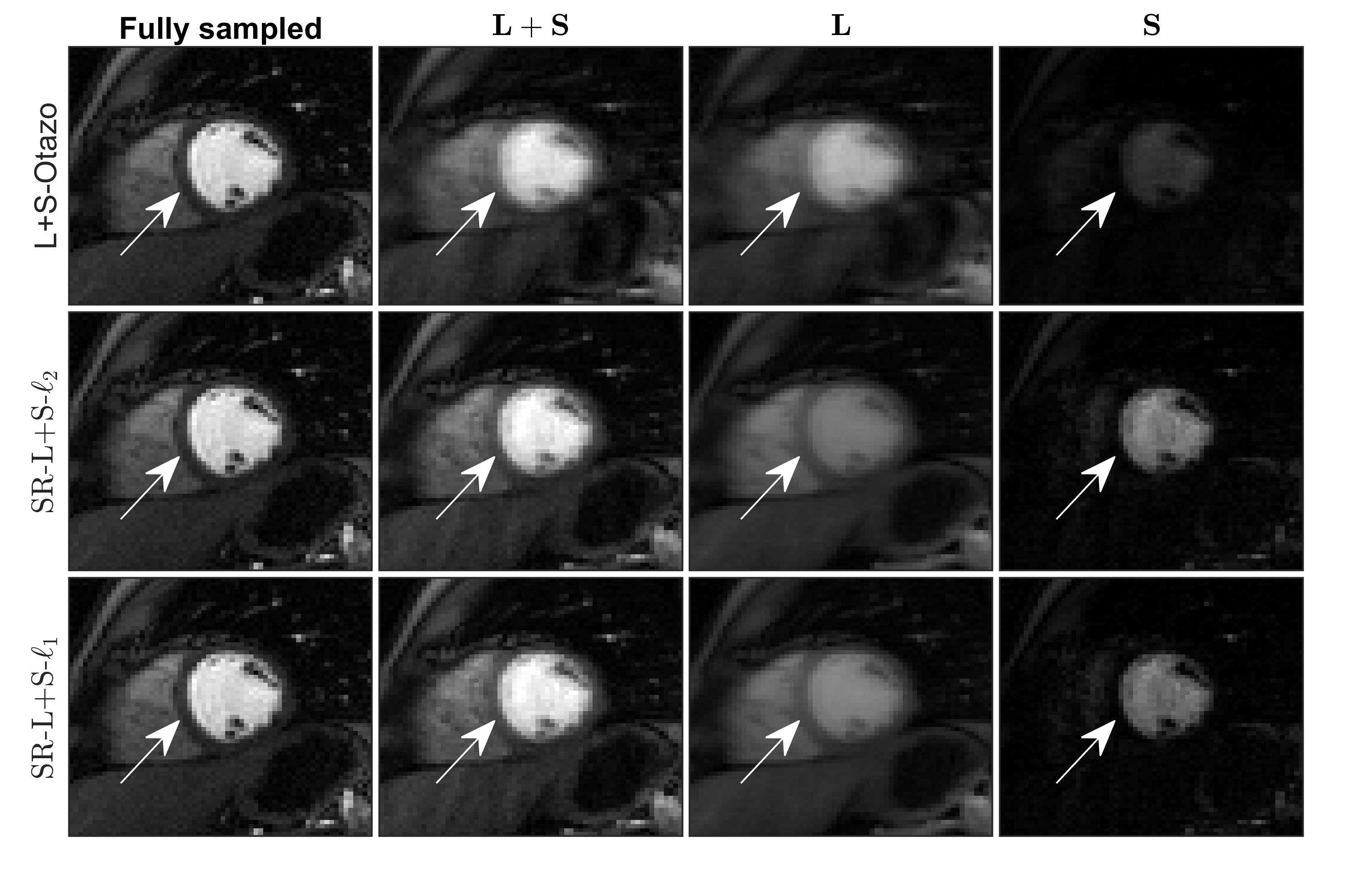}}
 \vspace{-0.1cm}
  \centerline{(a)}\medskip
\end{minipage}
\begin{minipage}[b]{1.0\linewidth}
  \centering
  \centerline{\includegraphics[width=8.5cm]{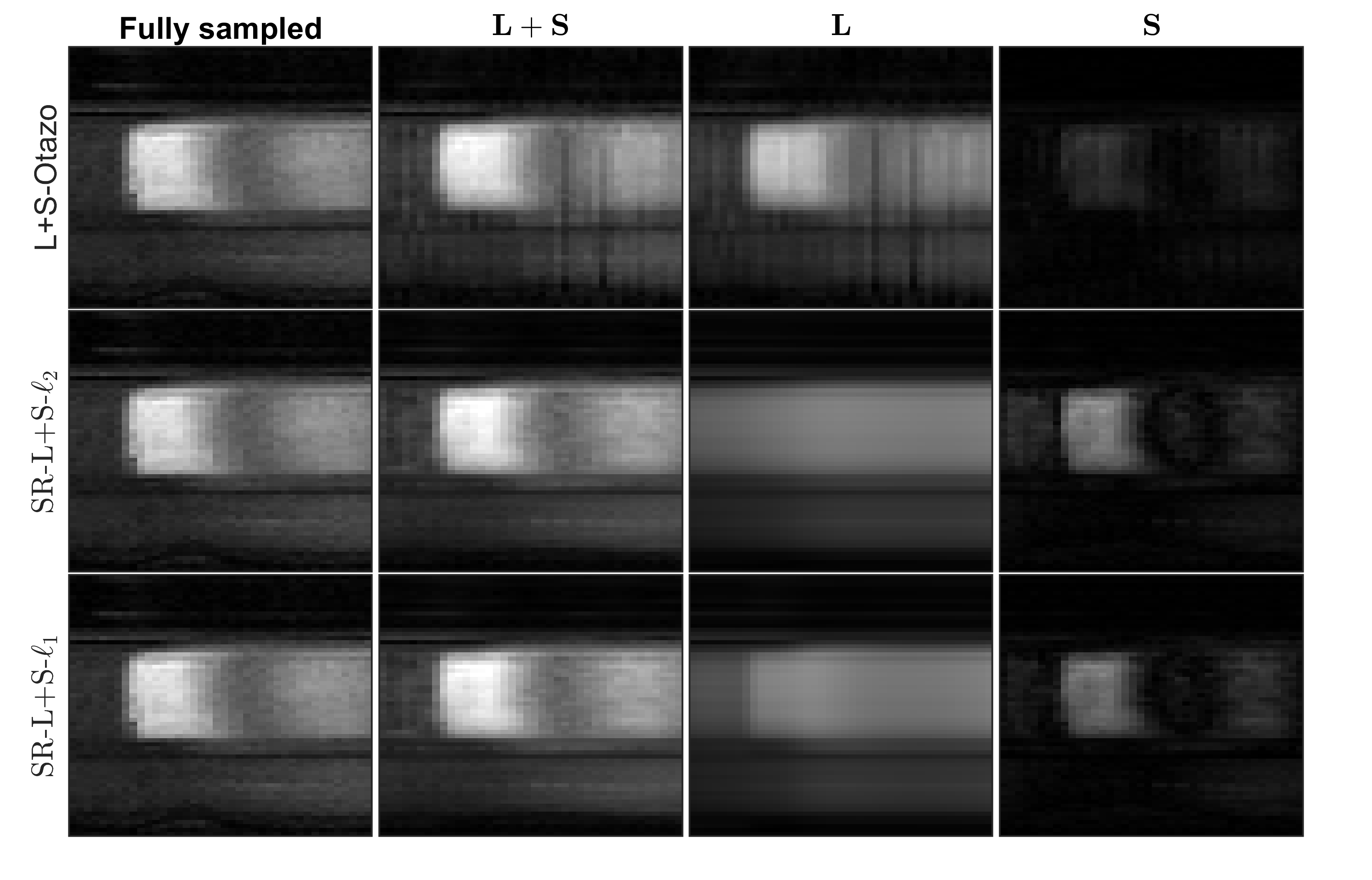}}
 \vspace{-0.1cm}
  \centerline{(b)}\medskip
\end{minipage}
\vspace{-0.9cm}
% \caption{The L and S component separation of cardiac perfusion dMRI sequence of a patient with coronary artery disease. (a) The x-y view of the original image (14th frame), the reconstructed image by L+S, and the reconstructed L component and S component using the SR-L+S-$\ell_1$. (b) The y-t views of reconstructed images by L+S and SR-L+S models.}
\caption{The L and S component separation of cardiac perfusion dMRI sequence of a patient with coronary artery disease. (a) The x-y view of the original image (14th frame), the reconstructed image by ${\bf X} = {\bf L}+{\bf S}$, and the reconstructed ${\bf L}$ component and ${\bf S}$ component using different L+S models. (b) The y-t views of reconstructed images by different models.}
\label{fig:fig3}
\end{figure}

\subsubsection{Separation of L and S components}

% In Fig.~\ref{fig:fig3}(a), we demonstrate SR-L+S-$\ell_1$ in decomposing the reconstructed image series into its constituent L (low-rank) and S (sparse) components. While acknowledging that achieving complete separation of background and dynamic components in cardiac perfusion remains a challenge \cite{otazo2015low}, our illustration in Fig.~\ref{fig:fig3}(a) highlights a notable observation. Specifically, within the sparse component S, a perfusion defect is obvious (indicated by an arrow), indicating successful background suppression in the low-rank L component.
% In Fig.~\ref{fig:fig3}(b), we visualize the reconstruction results for different time frames in the y-t domain, compared to their reference images. Our proposed approach, SR-L+S-$\ell_1$, demonstrates reduced temporal fluctuations compared to the L+S method \cite{otazo2015low}. Notably, while the S component exhibits sparsity, the L component evolves gradually over time and encapsulates the most correlated aspect (background) of the contrast enhancement in the Cardiac dataset.

Fig.~\ref{fig:fig3} shows the reconstruction results on the cardiac data by ${\bf X} = {\bf L} + {\bf S}$, along with its recovered background ${\bf L}$ and dynamic ${\bf S}$ components. From the x-y view (Fig.~\ref{fig:fig3}(a)), both the SR-L+S models provide a better visualization of the perfusion defect (as indicated by an arrow) in the sparse component ${\bf S}$ than the original L+S model, with the background successfully suppressed. More contrast is observed between the healthy part of the myocardium and the lesion. From the y-t view (Fig.~\ref{fig:fig3}(b)), we can clearly see that the SR-L+S models present higher temporal fidelity with respective to the original image sequence, with substantially reduced temporal blurring artifacts. It is noteworthy that the our method can recover ${\bf L}$ that  can better encapsulate the slow variation or temporal correlations of the background structure, while revealing more pronounced dynamic activities sparsely distributed in ${\bf S}$. In contrast, substantial dynamic activities are still present in the background part of the images for the original L+S model. This implies better separation of background and foreground components of dMRI by our method, owing to the additional smoothness regularization that can further smooth out the remaining noise and dynamic activities from the background part. Nevertheless, we observe that the use of $\ell_2$ smoothness prior in the SR-L+S model may in certain degree over-smooth the background part compared to the $\ell_1$ penalty.

\section{Conclusion}
We developed a novel framework based on SR-L+R decomposition to reconstruct dMRI from undersampled data. The SR-L+R is an extension of the L+S formulation, that applies joint low-rank and smooth regularization to fully exploit both the global and local temporal correlations in the slowly-varying background component, enhancing reconstruction robustness to noise.
% to make the reconstruction more robust to noise. 
An efficient algorithm based on the proximal gradient method is developed to solve a convex optimization for the SR-L+R decomposition. Experimental results on dynamic cardiac and synthetic data shows that the proposed method can provide better recovery of the dynamic MR images, and its background and dynamic components, compared to existing methods. Future work could extend the proposed model using tensor decomposition \cite{roohi2017multi} and deep unfolded network that utilizes deep networks to learn the regularization parameters and the proximal mappings in the iterative solver as in \cite{huang2021deep, liu2020deep}.

\bibliographystyle{unsrtnat}
\bibliography{references}  %%% Uncomment this line and comment out the ``thebibliography'' section below to use the external .bib file (using bibtex) .





% \end{thebibliography}

\end{document}